\documentclass[prc,preprint,superscriptaddress,nofootinbib]{revtex4}
\usepackage{amsfonts}
\usepackage{amssymb}
\usepackage{multirow}
 \usepackage{booktabs}
\usepackage{amsfonts}
\usepackage{amsmath}
\usepackage{amssymb}
\usepackage{amsthm}
\usepackage{epsf}
\usepackage[dvips]{graphicx}
\def\dfrac{\displaystyle\frac}

\def\bk{{\mbox{\boldmath$k$}}}

\def\bp{{\mbox{\boldmath$p$}}}

\def\bxi{\mbox{\boldmath{$\xi$}}}

\begin{document}

\title{Accounting for the analytical properties of the quark propagator from
Dyson-Schwinger equation}

\author {S.~M. Dorkin}
\affiliation{Bogoliubov Lab.~Theor.~Phys., 141980, JINR, Dubna,
 Russia}
\affiliation{International University Dubna, Dubna, Russia }

 \author{L.~P. Kaptari}
 \affiliation{Bogoliubov Lab.~Theor.~Phys., 141980, JINR, Dubna,
 Russia}
\affiliation{Helmholtz-Zentrum Dresden-Rossendorf, PF 510119, 01314
Dresden, Germany}

\author { B.~K\"ampfer}
\affiliation{Helmholtz-Zentrum Dresden-Rossendorf, PF 510119, 01314
Dresden, Germany}
\affiliation{Institut f\"ur Theoretische Physik, TU Dresden, 01062 Dresden, Germany}

\begin{abstract}
 An approach based on combined solutions of the
 Bethe-Salpeter (BS) and Dyson-Schwinger (DS) equations within the
 ladder-rainbow approximation
 in the presence of singularities  is proposed to describe the
 meson spectrum as quark-antiquark bound states.
 We consistently implement into the BS equation the quark propagator functions from the DS equation,
 with and without   pole-like singularities, and show that, by knowing
  the precise positions of the poles and their residues,
  one is able to develop reliable methods of obtaining finite interaction BS kernels and
  to solve   the BS equation numerically.
   We show that, for bound states with masses $M < 1$ GeV, there are no singularities in
   the propagator functions when employing the infrared part of the Maris-Tandy kernel
   in truncated BS-DS equations.
   For $M >1 $ GeV, however,  the propagator functions reveal
   pole-like structures.
   Consequently, for each type of mesons (unflavored, strange and charmed)
    we analyze  the relevant  intervals  of
    $M$  where the pole-like singularities
    of the corresponding quark propagator  influence the solution of
    the BS equation and develop
    a framework within which they can be consistently accounted for. The BS equation
    is solved for pseudo-scalar and vector mesons. Results are in a good agreement with experimental data.
   Our analysis is  directly related to the future physics programme at FAIR
   with respect to open charm degrees of freedom.

\end{abstract}

\maketitle
\section{Introduction}

 The investigation of mesons as bound states of quarks is of  fundamental interest
 for understanding the low-energy degrees of freedom of
 strong interaction and its relation to QCD.
 It is tightly connected with non perturbative methods in QCD and   directly related
 to study of such important phenomena as dynamical
  chiral symmetry breaking, confinement of quarks, mass splitting of meson multiplets etc.
 In principle, lattice QCD simulations can provide "experimental" information on most
 non-perturbative effects of QCD. However, with respect to some practical limitations
  in lattice calculations, it is extremely important to elaborate in parallel
   reliable phenomenological
 or semi-phenomenological approaches
 to describe the main features of non-perturbative QCD. Such approaches
  would allow one to extend calculations
 to large distances, nowadays  inaccessible for exact calculations.
   From the other side, the
 elaboration  of consistent models for QCD bound states in vacuum
 can serve as clue in understanding the in-medium properties of hadrons at high densities and temperatures.
For instance, the planned experiments at FAIR, GSI~\cite{CBM,PANDA} and NICA~\cite{NICA},
offer the next-generation investigations of charmed probes in proton and anti-proton
induced reactions at nuclei as well as heavy-ion collisions accessing  the maximum baryon
density region. In contrast, the running experiments at RHIC and LHC address,  among other
important issues, the behaviour of charmed probes in hot matter in the deconfinement
region. An ultimate prerequisite of the interpretation of current and future experiments
including high-statistic  charmed probes is the firm theoretical understanding of the meson spectrum in
such a mass range. Once this is accomplished one goes ahead towards  in medium-effect by appropriate methods.


  In the present paper, the mesonic bound states in vacuum are described
  within the framework of the homogeneous Bethe-Salpeter (BS) equation~\cite{nakanishi}
  with momentum dependent quark mass functions, determined by the Dyson-Schwinger (DS)
  equation. For the sake of consistency, in both, the BS and DS equations
  one  uses identical interaction kernels.
  In principle, if one were able to solve the complete set of  Dyson-Schwinger equations for
  quark and gluon propagators and vertex functions as well,
 the approach would  not depend on additional parameters. However,  due to known
 difficulties, in  real calculations
 one restricts oneself to   first, one-loop  term of the perturbative series and, based on the
 obtained results, establish a general form of the  phenomenological gluon propagators to be used in DS and BS equations.
  In elaborating such approaches   it is important that the suggested BS kernels, which implicitly is also
  contained in the DS equation, and   dressed quark-gluon vertices, pertaining both BS and DS equations,
  to be consistent to each  other, i.e. to guarantee at least
  the Ward-Takahshi identity~\cite{rob-11}. One of  such an approach is known as
  Maris-Tandy model~\cite{Maris:2003vk} which is based on  the   rainbow-ladder
  approximation of the DS equation, see also \cite{physRep}. The merit of the approach is that, once the effective parameters are fixed,
 the  spectrum of known mesons is supposed to  be described   on the same footing,
 including also excited states.
 The model has been applied
 to explain successfully many spectroscopic data, such as meson
 masses~\cite{rob-1,Maris:1999nt,Maris:2003vk,Holl:2004fr,Blank:2011ha,ThomasHilger},   electromagnetic properties
 of pseudoscalar mesons and their radial excitations~\cite{Krassnigg:2004if}
 and  other   observables~\cite{ourFB,wilson,tandy1,David,Alkofer,fisher,Roberts}.

The fact that the model encounters difficulties in describing heavy mesons, $M > 1$ GeV, with
at least one light (u, d or s)  quark   is not often mentioned. However, now this is not a
critical defect of the model since
the source of difficulties seems to be firmly understood - these are pole-like singularities
in the propagator functions from the  DS equation at large meson masses~\cite{ourPRCSingularities}.
An accurate treatment of the singularities relevant to the kinematical region of the BS equation
can essentially facilitate computations of the BS kernel  removing in such a way
the  difficulties and allowing to obtain stable results.

 In the present paper we continue our investigation~\cite{ourPRCSingularities} of
 the prerequisites to the interaction
 kernel of the combined Dyson-Schwinger and Bethe-Salpeter  formalisms to describe
 the meson mass spectrum including heavier mesons and excited states. In the previous paper~\cite{ourPRCSingularities}
  it
 was argued that the propagator functions from the DS equation
  are not analytical functions in the Euclidean
 complex plane impeding numerical solutions of the BS equation.   Our goal herein
 is to analyze the analytical properties of the quark propagators in the whole
 kinematical region relevant to light and semi-heavy meson masses,  supply  a method
 to implement the singular propagators in solving numerically the BS equation and solve
 the BS equation for mesons in the presence of singularities.
 To this end  we solve the
   DS equation in the rainbow ladder approximation by making use of the hyperspherical
   harmonics basis to decompose the propagators and the corresponding potential
   and solve numerically the resulting DS equations for the coefficients of such a
   decomposition. Then, the further  analysis of quark singularities
   is based on a combined application of  Rouch\'e's theorem and a graphical
   representation of the inverse propagators  as  vortex fields of the corresponding
   complex functions, see Ref.~\cite{ourPRCSingularities} for details. In the present paper we
   restrict ourselves
   to meson masses up to $M = 3.5$ GeV and, correspondingly, investigate
    all the relevant poles of
   quarks of different flavors  (u, d, s and c quarks) solely in this kinematical region.
   Then we suggest a method of accounting for singularities and  solve the BS equation
   for light and heavier mesons.

It should be noted that there exist other approaches based on the same physical ideas
of exploring effective quark interactions. For instance, in Ref.~\cite{dorokhov} (and references therein quoted)
 it was demonstrated that within an approach with instanton fluctuations of the QCD vacuum
it is possible to describe the mechanism of formation of mesons as bound states of quarks and
to analyse their main physical properties, including the relation between quark propagators
and quark condensates. Other approaches  employ simpler interactions,
e.g. a  separable interaction for the effective
coupling~\cite{David}. Such models  describe also fairly well  properties of light mesons,
nevertheless, the investigation of  heavier mesons and excited states,
consisting even of light (u, d and s)  quarks, requires implementations of
more accurate numerical methods to solve the corresponding equations.

 Our paper is organized as follows.
 In Secs.~\ref{s:bse},~\ref{ss:hyper}, ~\ref{ss:numerical} and~\ref{s:DS}
 we briefly discuss the truncated
  BS and DS equations within the rainbow approximation
   relevant to describe the mesons as quark-antiquark
 bound states. The domain of the complex plane of
 Euclidean momenta, where the solutions of the BS equation are sought, and the corresponding
 propagator functions
 are specified in Subsecs.~\ref{domain} and~\ref{explic}.
 Subections~\ref{right}--
\ref{poles} are
  dedicated to the solution of the truncated DS equation for complex  momenta and to a thorough
  analysis of the singularities of the propagator functions in the domain of
  the Euclidean space relevant to BS solutions for  mesons with masses up to $3.5$ GeV.
  Convenient parametrizations for the propagator functions of u, d, s and c quarks used to
  determine the BS kernel in the domain of its analyticity are presented in  Subsec.~\ref{parametriz}.
   The subsequent Sections are aimed at investigations of
  methods of solving the BS equation in presence of singularities.
 In Subsec.~\ref{ps:mesons} we present results of numerical calculations of the
 mass spectra of the pseudo-scalar mesons. Alternative numerical approaches and
 nuances in solving the BS equation for heavy mesons with quarks of equal masses are
 briefly discussed in Subsecs.~\ref{excited},~\ref{exhaust} and~\ref{equalMass}.
 Summary and conclusions are collected in Sec.~\ref{summary}. In the Appendix some useful relations
 used in solving the BS equation are presented.

\section{Bethe-Salpeter equation}
\label{s:bse}
\subsection{Ladder-rainbow approximation}
To determine the bound-state energy (mass) of a quark-antiquark pair one needs to solve the Bethe-Salpeter equation, which
in the ladder approximation  (hereafter referred to as truncated Bethe-Salpeter (tBS) equation)  and
in Euclidean space,  reads
\begin{eqnarray}
    \Gamma(P,p) =   -\frac 43  \int \frac {d^4k}{(2\pi)^4}
    \gamma_{\mu}S\left(\frac12 P+k\right) \Gamma(P,k) S\left(-\frac12  P+k\right)\gamma_{\nu}
    \left [g^2    {\cal D}_{\mu \nu}(p-k)\right ] \: ,
\label{bse}
\end{eqnarray}
where the interaction kernel $g^2 {\cal D}_{\mu \nu}(k^2)\equiv {\cal D}(k^2) d_{\mu\nu}(k^2)$ is
chosen in the Landau gauge,
$  d_{\mu\nu}(k^2)   = \left[ \delta_{\mu\nu}-k_\mu k_\nu/k^2 \right]$,
 $\Gamma(P,p)$ is  the tBS vertex function,  and
$S(k_{1,2})= -i\gamma\cdot k_{1,2}\sigma_v(k_{1,2}^2) +\sigma_s(k_{1,2}^2)  $
 are the  quark propagators   with the propagator functions $\sigma_{v,s}(k^2)$.
 The total $P$ and relative momenta $p$ are defined as $P=p_1+p_2$ and  $  p=(p_1-p_2)/2$,
 where $p_{1,2}$ are the momenta of the constituent quarks.
 The two quarks  interact via gluon exchange encoded in
$\left [g^2    {\cal D}_{\mu \nu}(p-k)\right ]$.
The vertex
function $\Gamma (P,k)$ is a $4\times 4$ matrix and, therefore, may contain
16 different functions.

 The interaction kernel is chosen within
   the
  ladder rainbow approximation.
  It  has been   widely used  to study the physics of dynamical chiral symmetry breaking
 \cite{Maris:1999nt}, decay
 constants~\cite{Maris:2003vk,Holl:2004fr,Blank:2011ha,Krassnigg:2004if} and other   observables
 \cite{ourFB} and has been found to
 provide a good agreement with experimental data.
 The employed vertex-gluon kernel in the rainbow approximation
   is chosen here in the form  \cite{Maris:1999nt,Maris:2003vk,Alkofer,Roberts}
\begin{eqnarray}
  {\cal D} (k^2) =
    \left(
        \frac{4\pi^2 D k^2}{\omega^2} e^{-k^2/\omega^2}
        + \cdots
    \right)
    \ ,
\label{phenvf}
\end{eqnarray}
 where the first term originates from  the effective infrared (IR) part of the interaction
 determined by soft non-pertubative effects, while  further ones (hidden in $\cdots$)
  ensure the correct ultraviolet (UV) asymptotic behaviour of the QCD running coupling.  A detailed investigation of
 the interplay of IR and UV  terms
 has shown~\cite{souglasInfrared,Blank:2011ha} that the IR part  is dominant for light  u, d and s quarks,
  with a decreasing role for heavier quark  masses (c and b) for which the
UV part may play a role  in forming  meson masses with $M > 4$ GeV. In the present
paper we consider mesons with masses up to $3.5$ GeV and, consequently, in the
interaction kernel ${\cal D}(k^2)$  the spelled out IR term is accounted for. It depends on two
parameter, $D$ and $\omega$. Since within the ladder approximation the tBS amplitude does not depend  on the total momentum $P$, in what follows in $\Gamma (P,p)$ and in all subsequent
partial amplitudes we omit $P$ as redundant notation.

The general structure of   vertex functions
describing bound states of spinor particles  has been
investigated in detail, for example, in \cite{ourUmn,dorByer,lastFischer}. To release from the matrix structure, the vertex
function  $\Gamma$ is  expanded into functions which in turn are determined by angular momentum and parity of the corresponding
meson known as the spin-angular harmonics~\cite{dorByer,ourFB,ourFB1}:
\begin{eqnarray}
    \Gamma(p)&=&\sum\limits_\alpha g_\alpha(p) \,{\cal
    T}_\alpha(\bp), \qquad
g_\alpha(p)= \int d\Omega_\bp\, \mathrm{Tr}\,[{\Gamma}(p){\cal
T}_\alpha^+(\bp)]\label{spex}.
\end{eqnarray}
The complete set of spin-angular matrices
 are chosen as: \\ (i) in the $^1S_0$ pseudo-scalar channel
\begin{eqnarray}
{\cal T}_1(\bp)=\frac{1}{\sqrt{16\pi}}\gamma_5,\qquad &&
{\cal T}_2(\bp)=\frac{1}{\sqrt{16\pi}}\gamma_0\gamma_5,\nonumber \\
{\cal T}_3(\bp)=-\frac{1}{\sqrt{16\pi}}\widehat n_{\bf p}\gamma_0\gamma_5,\qquad &&
{\cal T}_4(\bp)=-\frac{1}{\sqrt{16\pi}}\widehat n_{\bf p}\gamma_5,
\label{nharms}
\end{eqnarray}
and  (ii) for the $^3S_1$--$^3D_1$ vector channel
 \begin{eqnarray}\begin{array}{ll}
{\cal T}_1(\bp) =\sqrt{\dfrac{1}{16\pi} }\widehat\xi_{\cal M}, 
&
{\cal T}_2(\bp)=-\sqrt{\dfrac{1}{16\pi}}\,\gamma_0\,\widehat\xi_{\cal M}, \\[10pt]
{\cal T}_3(\bp)=-\sqrt{\dfrac{3}{16\pi}} (n_{\bf p} \xi_{\cal M}), 
&{\cal T}_4(\bp)=\sqrt{\dfrac{3}{32\pi}}\, {\gamma_0}\left[-(n_{\bf p} \xi_{\cal M})+
\hat n_{\bf p} \widehat \xi_{\cal M}  \right],\\[10pt]    \label{nharmd}
{\cal T}_5(\bp)=\sqrt{\dfrac{1}{32\pi}}\left[\widehat\xi_{\cal M}+ {3}\,
(n_{\bf p} \xi_{\cal M}) \widehat n_{\bf p} \right], 
&{\cal T}_6(\bp)=\sqrt{\dfrac{1}{32\pi}}\gamma_0\left[\widehat\xi_{\cal M}+ {3}\,
(n_{\bf p} \xi_{\cal M}) \widehat n_{\bf p} \right],\\[10pt]
{\cal T}_7(\bp)= -\sqrt{\dfrac{3}{16\pi}}\,\gamma_0 (n_{\bf p} \xi_{\cal M}) ,\quad
&{\cal T}_8(\bp)= \sqrt{\dfrac{3}{32\pi}}\,\left[-(n_{\bf p} \xi_{\cal M})+
\hat n_{\bf p} \widehat \xi_{\cal M}  \right],
\end{array}
\end{eqnarray}
where all the above  scalar products are written in Minkowski space and the unit vector $n_{\bf p}$
is defined  as $n_{\bf p}=(0,{\bf p}/|{\bf p}|)$.
The left hand side of (\ref{nharmd}) depends implicitly on $\mathcal{M}$, which denotes the components
of the polarization vector $\xi_{\cal M}\equiv (0,\bxi_{\cal M})$ fixed by
$\bxi_{+ 1} =-(1,i,0)/\sqrt{2}$, $\bxi_{- 1} =(1,-i,0)/\sqrt{2}$,
$\bxi_{0} =(0,0,1)$.
Similar complete sets of spin-angular harmonics  have been employed also
in Refs.~\cite{Alkofer,lastFischer}.

With Eqs.~(\ref{spex})-(\ref{nharmd}) the integral matrix form of the  BS equation
(\ref{bse}) can be reduced to a system of  two-dimensional integral equations with respect to
the partial vertices  $g_\alpha(p)$, cf.~\cite{dorByer,ourFB1}.

 \subsection{The hyperspherical decomposition\label{hyper} }
\label{ss:hyper}
 To further reduce the dimension of the integral
 we   decompose, in Euclidean space, the partial vertices $g_\alpha(p)$ and
 the interaction kernel  ${\cal D} (p-k)$  in (\ref{bse})
  over the basis of spherical harmonics
 $ {\rm Y_{lm}}(\theta,\phi)$ and
 normalized Gegenbauer polynomials $X_{nl} (\chi)$, i.e. we use the
 hyperharmonic   basis

\begin{eqnarray}
Z_{nlm}(\chi) = X_{nl} (\chi) {\rm Y_{lm}}(\theta,\phi)\equiv \sqrt{\frac{2^{2l+1}}{\pi} \frac
{(n+1)(n-l)!l!^2}{(n+l+1)!}} \sin^l\chi G_{n-l}^{l+1}(\cos \chi) {\rm Y_{lm}}(\theta,\phi),
\label{hsg}
\end{eqnarray}
  where $G_{n-l}^{l+1}(\cos \chi)$ are
the  Gegenbauer polynomials of the hyperangle $\chi$ with
$\cos \chi =\dfrac{p_4}{\tilde p}$ and
$\sin\chi =\dfrac{|{\bf p}|}{\tilde p}$
and $\tilde p=\sqrt{p_4^2+\bp^2}$   of an Euclidean 4-vector  $p$.
 Then, the partial decomposition of the vertex functions $g_\alpha (p_4,\bp)$ ($\alpha =1\ldots 4$ and
 $\alpha =1\ldots 8$
 for pseudoscalar and vector mesons, respectively)   and interaction kernel
 reads
\begin{eqnarray}
\label{exp11}&&
g_\alpha (p_4,\bp)=\sum\limits_{n l_\alpha} \varphi_{\alpha,l_\alpha}^n
(\tilde p)\,X_{nl_\alpha}(\chi_p){\cal T}_\alpha(\bp),
\label{er} \\ &&
%
{\cal D}( p-k ) =2\pi^2 \sum_{\kappa\lambda \mu}
\frac{1}{\kappa+1} V_\kappa(\tilde p,\tilde k)X_{\kappa\lambda }(\chi_p) X_{\kappa\lambda }(\chi_k) Y_{\lambda \mu}(\Omega_p)Y_{\lambda \mu}^*(\Omega_k). \label{exp12}
\end{eqnarray}
We label here and in the following    the modulus of an Euclidean vector
$p=(p_4,\bp)$ by a  tilde, i.e.
$\tilde p\equiv \sqrt{p_4^2+\bp^2}$.
Actually, in Eq.~(\ref{exp11}) the summation over $l_\alpha$  is restricted by
 the corresponding orbital momentum
    encoded in the spin-angular matrices  ${\cal T}_\alpha(\bp)$.
It can be seen from Eq. (\ref{nharms}) that  the spin-angular harmonics
${\cal T}_{1,2}(\bp)$ in the $^1S_0$ channel (pseudoscalar mesons) carry the momentum  $l_\alpha=0$,
 while for ${\cal T}_{3,4}(\bp)$ one has $l_\alpha=1$.
Analogously  for vector mesons (cf.  Eq.~(\ref{nharmd})) $l_\alpha=0$ in ${\cal T}_{1,2}(\bp)$,
$l_\alpha=2$ in ${\cal T}_{5,6}(\bp)$ and $l_\alpha=1$ otherwise.

Changing the integration variables to the hyperspace,
$d^4k =\tilde k^3\sin^2\chi_k \sin\theta_k d \tilde k d\chi_k  d\theta_k d\phi_k$,
inserting (\ref{exp12}) and (\ref{er}) into
 (\ref{bse}) and performing the necessary angular integrations we
obtain  a system of   integral equations for
the expansion coefficients~$\varphi_{\alpha,l_{\alpha}}^n$:
\begin{eqnarray}
\varphi_{\alpha,l_\alpha}^n(\tilde p) =
\sum_\beta \sum\limits_{m=1}^{\infty} \int  d\tilde k \tilde{k}^3  S_{\alpha \beta}(\tilde p,\tilde k,m,n)
  \varphi_{\beta,l_\beta}^m(\tilde k)
\label{eqnf}.
\end{eqnarray}

The explicit expressions for the coefficients $S_{\alpha \beta}(\tilde p,\tilde k,m) $
result from the corresponding angular integrations over $d\Omega_{\bf p}$, $d\Omega_{\bf k}$,
$\sin^2 \chi_k d\chi_k$ and $\sin^2 \chi_p d\chi_p$ which
can be written in the form
\begin{eqnarray}&&
S_{\alpha \beta}(\tilde p,\tilde k,m,n) = \sum_{\kappa}\int\sin^2\chi_k d\chi_k X_{ml_\beta}(\chi_k)
X_{\kappa \lambda}(\chi_k)\sigma_{s,v}(  k_1^2)\sigma_{s,v}(  k_2^2) A_{\alpha\beta}(\tilde p,\tilde k,\kappa,\chi_k,n),
\label{kernel}
\end{eqnarray}
where
\begin{eqnarray}
   k_{1,2}^2 = \left (\frac12 P\pm k \right)^2=-\displaystyle\frac{1}{4}M^2 + \tilde k^2 \pm i M \tilde k \cos\chi_k,
  \label{parequation}
  \end{eqnarray}
with $k$ as the  relative momentum of two quarks, and the total momentum $P=(iM,{\bf 0})$.
The quantity   $ A_{\alpha\beta}$  resulting from
evaluations  of traces and angular integration in the 3-momentum space,  has schematically  the  form
\begin{eqnarray}&&
 A_{\alpha\beta}(\tilde p,\tilde k,\kappa,\chi_k)\simeq
  \int d\Omega_\bp d\Omega_\bk \sin^2\chi_p d\chi_p   V_{\kappa}(\tilde p,\tilde k)
  d_{\mu\nu}\left((p-k)^2 \right) \times
\nonumber\\ &&X_{nl_\alpha}(\chi_p)
  X_{\kappa\lambda}(\chi_p) {\rm Y}_{\lambda\mu}(\theta_p,\phi_p){\rm Y}_{\lambda\mu}^*(\theta_k,\phi_k)
Tr \left[\gamma_\mu ...{\cal T}_\beta(\bk)...{\cal T}_\alpha^+(\bp) \gamma_\mu \right].
\label{Aab}
\end{eqnarray}
The angular structure of the integrand in~(\ref{Aab})  is rather simple: it contains
a series of products of spherical harmonics, Gegenbauer polynomials and
scalar products of $(\bp \bk)\sim {\rm Y^*_{1m}}(\theta_p,\phi_p){\rm Y_{1m}}(\theta_k,\phi_k)$, i.e. all angular  integrations over $d\Omega_\bp d\Omega_\bk d\chi_p$ can be performed explicitly. These integrations  provide a smooth
quantity $A_{\alpha\beta}(\tilde p,\tilde k,\kappa,\chi_k)$ being free of any singularities. That  means
 the analytical structure of the kernel (\ref{kernel}) is entirely determined by the propagator functions
$\sigma_{s,v}(\tilde k^2)$.

Recall  that in Eqs. (\ref{eqnf})-(\ref{Aab}) the indices  $(\alpha,\beta)$   label the
tBS components in the spinor space
($\alpha,\beta=1\ldots 4$ for   pseudoscalar mesons and $\alpha,\beta=1\ldots 8$
for vector mesons), $(m,n)$ denote the number of terms in the Gegenbauer
decomposition~(\ref{er}),
and $l_{\alpha,\beta}$ are entirely determined by the corresponding components
${\cal T}_{\alpha,\beta}$
of the spin-angular basis.

\subsection{Numerical solutions }
\label{ss:numerical}
Now we proceed with solving    the BS equation (\ref{eqnf})  for the partial vertices
$\varphi_{\alpha,l_\alpha}^n(\tilde p)$.
Equations (\ref{eqnf})-(\ref{Aab})  represent the desired system
   of the BS one-dimensional integral equations of the   Fredholm type within the hyperspherical
   harmonics formalism to be solved numerically.
   Before choosing a specific computational
   algorithm one has to analyze at least the
  existence and uniqueness of the solution.  Obviously, this issue is directly connected
   to the properties of the interaction kernels.
  As known,  the main   requirement for the existence of solutions of  Fredholm type
     equations is the finiteness of the integral kernel, i.e. of  the
     quantity  $S_{\alpha \beta}(\tilde p,\tilde k,m,n)$  in our case. As  mentioned above
the function $A_{\alpha\beta}(\tilde p,\tilde k,\kappa,\chi_k)$,
which determines the kernel $S_{\alpha \beta}(\tilde p,\tilde k,m,n)$, is finite by construction.
 That means the only source of troubles can appear from  the propagator functions $\sigma_{s,v}(\tilde k^2)$  in (\ref{kernel}),
which, in turn, are the solution of the tDS equation with the same interaction kernel (\ref{phenvf}).
Accordingly, prior to solve the tBS equation, firstly we shall proceed with an analysis
of the propagator functions $\sigma_{s,v}(\tilde k^2)$  from the Dyson-Schwinger equation in Euclidean space.
\section{Dyson-Schwinger Equation}
\label{s:DS}
\subsection{The relevant region for the truncated Dyson-Schwinger
  equation}\label{domain}
  We are interested
  in the analytical structure of the propagator functions $\sigma_{s,v}(\tilde k^2)$  inside and
  in the  neighbourhood of the complex momentum region in the  Euclidean space dictated by the tBS equation (\ref{bse}). This
  momentum region is displayed as the dependence of the imaginary part of the quark momentum squared Im\ $\tilde k^2$  on
  its real part
  Re\  $\tilde k^2$ from    the tBS equation
   determining
  in  Euclidean complex momentum plane a domain restricted by  a parabola
  \begin{equation}
  Im~\tilde k^2=\pm~M\sqrt{{\rm Re}\,\tilde k^2~+~\frac{1}{4}M^2} \label{parabola}
  \end{equation}
   with  vertex at
 Im\ $\tilde k^2=0$ at Re\ $\tilde k^2 =-M^2/4 $
  depending  on the meson mass $M$.

  Note that regardless of the form of the interaction kernel, the
   investigation  of the analytical structure of the quark propagator
  is of  great importance, if the propagators exhibit singularities within the corresponding
  parabola, thus hampering the numerical procedure of solving the tBS equation.
  On the other side, the knowledge of the nature of singularities and their exact location
 in the complex plane will allow one to develop   effective  algorithms adequate for
 numerical calculations. For instance, if one determines exactly the
 domain of analyticity of the propagator functions, one can take advantage of the fact that
any  analytical function can always be approximated by rational
 complex  functions \cite{walsh}.  Then, one can   parametrize the
 integrand  in the tBS equation  by simple functions which will allow  to carry out some
 integrations analytically. Such parametrisations have been suggested in Ref.~\cite{souchlas} for
 meson masses $M <  1$ GeV  for which  the propagator functions  have been  found
  to be analytical.
Unfortunately, as shown in Ref.~\cite{ourPRCSingularities}, for larger meson masses the propagator functions
exhibit singularities within the domain of tBS integration and, as a consequence,
parametrizations by rational functions
are not possible.   Nevertheless, even  in this case, if the propagator  functions have only isolated poles
 with  known locations and  residues, then calculations can be significantly simplified by splitting the singular  functions
 into two terms, one being analytical in the considered region and the other one having
  a simple pole structure, as discussed below.


  In  Euclidean space the quark propagator obeys    the  truncated Dyson-Schwinger  equation
\begin{eqnarray}
S^{-1}(p)= S_0^{-1}(p) + \frac 43 \int \frac
{d^4 k }{(2\pi)^4} \left[g^2 {\cal D}_{\mu \nu}(p-k) \right]\gamma_{\mu} S(k)
\gamma_{\nu}\: ,
\label{sde}
\end{eqnarray}
where $S_0^{-1}=i \gamma \cdot p + m_q$ and ${m}_q$ is the bare current  quark mass. To emphasize
the replacement of combined gluon propagator and vertex we use, as in Eq.~(\ref{bse}), the notation
$[g^2D_{\mu\nu}]$, where an additional power of $g$ from the second undressed vertex is
included.  For a consistent
treatment  of  dressed quarks and their bound states, the dressed gluon propagator
$[g^2 {\cal D}_{\mu \nu}(p-k)]$ must
be   the same in the tBS  (\ref{bse})   and the tDS equations  (\ref{sde}).

\subsection{Propagator functions}\label{explic}
The tDS equation~(\ref{sde})    is a four dimensional integral equation
 in matrix form. Simplifications can be achieved in exactly the same manner as for the BS equation
 using the same spin matrices and the hyperspherical harmonics basis.
Recall that the calculation of the renormalized Feynman diagrams leads to a
fermion propagator depending  on two  functions, e.g.
the renormalization constant $Z_2$ and the self-energy $\Sigma(p)$.
Instead of $Z_2$ and  $\Sigma(p)$ one can introduce other two quantities $A(p)$ and $B(p)$ or,
  alternatively, $\sigma_s(p)$ and $\sigma_v(p)$.
In terms of these functions the  dressed quark propagator $ S(p)$ reads~\cite{Roberts:2007ji,Maris:2003vk}
\begin{eqnarray}&&
    S^{-1}(p)= i \gamma \cdot p A(p)+ B(p),  \quad \quad S(p) =
-i\gamma \cdot p \sigma_v(p) +\sigma_s(p) \: ,\label{quarkpr1}
\end{eqnarray}
with
\begin{eqnarray}
\sigma_v(p)=\frac{A(p)}{p^2A(p)^2+B(p)^2}, \quad
\sigma_s(p)=\frac{B(p)}{p^2A(p)^2+B(p)^2}.
    \label{quarkpr}
\end{eqnarray}

The resulting system of equations to be solved  is
  a system of one-dimensional integral equations with respect to
  $A(p)$ and $B(p)$ (for details, see \cite{ourPRCSingularities,ourFB}), which we solve numerically. Independent parameters are
$\omega$, $D$ and $m_q$. We find that the iteration
procedure   converges  rather fast and  practically does not depend  on
the choice of the trial start functions for $A(p)$ and $B(p)$.

In our subsequent analysis we
employ the effective parameters
from Refs.~\cite{Alkofer,Roberts}, $\omega=0.5$ GeV and $D=16 \, {\rm GeV}^{-2}$.

\subsection{Solution in the right hemisphere, ${\mbox{\boldmath$ Re\ \tilde p^2 >0$}} $ }
\label{right}

 Evidently, the functions $A(p)$ and $B(p)$  are real at real values of $\tilde  p^2$. However, since in the
 tBS equation
 the domain of definition of the propagator functions is complex, one needs an analytical continuation of the solution
 from the real  positive axis $Re~\tilde  p^2>0$ to the whole complex domain inside the corresponding parabola.
In our calculations the parabolic integration domain for solving
 the tBS equation we conveniently divide  into two parts: (i) one
(infinite) region where Re~$\tilde  p^2 > 0$, and (ii) a second one where Re~$\tilde  p^2 < 0$, which is
restricted  by the meson mass $M$, i.e. with the minimum (negative) value
Re~$\tilde  p^2 =- {M^2}/{4}$.

From   the tDS equation~(\ref{sde}) it is explicitly seen that, in the right
hemisphere,  the integrals converge in the tDS equation.
That means, an analysis of the behaviour of the solution for Re~$\tilde p^2 > 0$ for large $|\tilde p^2 |$
can be accomplished  directly
by means of the real solutions $A(p)$ and $B(p)$ obtained along the real axis in the tDS equation
and by  utilization subsequently   the  tDS equation with complex external momenta
to compute   $A(p)$ and $B(p)$   in the desired domain.

 Another method of finding $A(\tilde p^2)$, $B(\tilde p^2)$, $\sigma_s(\tilde p^2)$ and $\sigma_v(\tilde p^2)$
 in the complex plane is to
 solve the tDS equation once along a closed contour and then to use the Cauchy theorem
 \begin{eqnarray}
 A(z) = \frac{1}{2\pi i}\oint  \frac {A(\xi)}{\xi-z} \  d\xi
\label{cauchy}
\end{eqnarray} to calculate the required quantities in
 any point inside the contour, see also see  Ref.~\cite{fisher}.

  Then, by the same method one can compute the Cauchy integrals of $A(\tilde p^2)$,
   $B(\tilde p^2)$, $\sigma_s(\tilde p^2)$ and $\sigma_v(\tilde p^2)$
  to check them for analyticity. We performed such an analysis and found that in the right hemisphere
 the Cauchy integrals for the  solutions $A(\tilde p^2)$, $B(\tilde p^2)$, $\sigma_s(\tilde p^2)$
 and $\sigma_v(\tilde p^2)$ for each
 type of quarks (u, d, s and c) is zero, i.e. they are analytical functions of $\tilde p^2$.
 Consequently, the integral kernel $S_{\alpha \beta}(\tilde p,\tilde k,m,n)$ in
 Eq. (\ref{kernel})   is finite in this case.
According to the theory of Fredholm type equations, one
 can infer that  such a kernel possesses a discrete spectrum of eigenvalues.

\subsection{Parametrization of propagator functions at ${\mbox{\boldmath$ Re\ \tilde p^2 >0$}} $ }
\label{parametriz}

 The interaction  kernel and its eigenvalues can be found numerically by solving  the tDS equation for
   $\sigma_{s,v}$ and by implementing this solution into the numerical procedure for tBS equation.
  In principle, one can  avoid such  cumbersome  calculations of $S_{\alpha \beta}(\tilde p,\tilde k,m,n)$
   by taking advantage of the analyticity  of $\sigma_{s,v}$ in the right hemisphere.  It is known~\cite{walsh} that
  for any analytical function one can find convenient parametrizations
 in terms of rational functions,  which  in turn   can be chosen in such a  form as to be able
 to calculate  the hyperangular integrals explicitly.

A convenient  choice  for the parametrization to calculate the integrals
over Euclidean momentum $k$  in Eq. (\ref{kernel}) could be of the following form
\begin{eqnarray}&&
\sigma_{s,v}(\tilde k^2)=\sum_i \frac{\alpha_i(s,v)}{\tilde k^2+\beta_i^2(s,v)} +
 \sum_i \frac{\alpha_i^*(s,v)}{\tilde k^2+\beta_i^{*2}(s,v)},
\label{parx}
\end{eqnarray}
where the complex parameters $\alpha_i$ and $\beta_i$ can be easily  obtained by fitting the corresponding
 solution along the real axis of $\tilde k^2$.  We use the  Levenberg–-Marquardt algorithm  for fitting.
 With such a monopole form of the parametrization the corresponding integral in (\ref{kernel}) can be reduced to
a sum of  integrals of the form
 \begin{eqnarray}
 {\cal I}^\lambda_{mn}(z) = \int\limits_{-1}^1 d   \xi (1-\xi^2)^{\lambda-1/2}\frac{G_{n}^\lambda(\xi)G_{m}^\lambda(\xi)
 }{\xi -iz}; \qquad {\rm\,\, with}\quad z\sim\dfrac{Re~\tilde k_{1,2}^2+\beta^2}{Mk},
 \label{base}
 \end{eqnarray}
 which  can be calculated explicitly (see Appendix, Eq.~(\ref{a2})).

We find that for each function in Eq.~(\ref{parx}) the first three terms,
which involve  12 parameters,
are quite sufficient to obtain a good approximation of the solution.
In Tables~\ref{tb1} and \ref{tb2} we present the sets of parameters $\alpha_{i}(s,v)$
and $\beta_{i}(s,v)$ for u/d, s and c quarks,
obtained for $\sigma_s$ and $\sigma_v$, respectively,
from a fit in the interval $-0.15\, {\rm( GeV/c)} ^2< \tilde k^2 < 10  {\rm (GeV/c)}^2$.
In spite of the quality of the  excellent fit
it should be   noted, however,   that  since the  parametrized
functions are of a rather simple shape, the  obtained sets of   parameters $\{ \alpha_i\}$ and $\{ \beta_i\}$
are far from being  unique, i.e.  one can achieve a similar quality of the fit with many other choices
of $\{ \alpha_i\}$ and $\{ \beta_i\}$. The only restriction is
that the "mass" parameters   $\{ \beta_i\}$ must not provide singularities, neither along the real axis, nor in the
complex plane inside the parabola, see also Ref.~\cite{Bhagwat:2002tx}.
The obtained  sets of parameters, Tables \ref{tb1} and \ref{tb2},
have been used in our calculations of the propagator functions at
$Re\   \tilde k^2 > 0$, where they are always analytical.

 \begin{table}[!ht]
 \caption{ The parameters
 $\alpha_i(v)$ and $\alpha_i(s)$   for the effective parametrizations, Eq.~(\ref{parx}), \\
 \hspace*{1.7cm} for  three values of the bare quark mass.
 }
    \begin{tabular}{c clll} \hline
    $m_q$ [MeV] & $\alpha_i$& \hspace*{15mm}  1 &\hspace*{15mm}   2&\hspace*{15mm}  3  \\ \hline
    \multirow{2}{30pt}{5}         & $\alpha_v$  \hspace*{2mm}     &(0.1915,\ 0.94311) & (0.15360,\ -0.18621)  &(0.15360,\  -0.18621)\\
                                  & $\alpha_s$[GeV] \hspace*{2mm} & (0.03018,\ 0.48657)&(-0.042732,\ -0.77654)&(0.015015,\ 0.02499)\\ \hline
     \multirow{2}{30pt}{115}      & $\alpha_v$\hspace*{2mm}  & (0.18503,\ 0.18389)&(0.21166,\ -0.74513) &(0.10244,\ 0.22828)\\
                                  & $\alpha_s$[GeV] \hspace*{2mm}  &(-0.20626,\ 0.59157)&(0.31850,\ 0.025363)& (-0.055932,\ 0.085008)\\  \hline
   \multirow{2}{30pt}{1000}       & $\alpha_v$\hspace*{2mm}        &( -0.17206,\ -0.77962)&(0.19591,\  -1.0314)&(0.48176,\ 0.51818)\\
                                  & $\alpha_s$[GeV]\hspace*{2mm}  &(1.3803,\ 0.94714) &(-0.94419,\  -0.27285)&(0.064465,0.071692)\\  \hline

    \bottomrule \end{tabular}
\label{tb1}
\end{table}

\begin{table}[!ht]
\caption{The parameters
 $\beta_i(v)$  and $\beta_i(s)$    for the effective parametrizations, Eq.~(\ref{parx}),\\
 \hspace*{1.7cm} for  three values of the bare quark mass.  }
    \begin{tabular}{c clll} \hline
    $m_q$ [MeV]                   & $\beta_i$[GeV]                &\hspace*{15mm} 1                &    \hspace*{15mm}              2&  \hspace*{15mm}            3  \\ \hline
    \multirow{2}{30pt}{5}         & $\beta_v$  \hspace*{2mm} & (0.5483,\ 0.19010)&(1.1385,\ 0.26394)&(1.1385,\ 0.26394)\\
                                  & $\beta_s$  \hspace*{2mm} &(0.53454,\ 0.17611)&(1.4018,\ 0.046165)&(- 1.1163, \ 1.2918)\\ \hline
     \multirow{2}{30pt}{115}      & $\beta_v$  \hspace*{2mm} & (0.62143,\ 0.23396)&(1.1353,\ 0.19534)& (0.69444,0.47076)\\
                                  & $\beta_s$  \hspace*{2mm} &(0.78697,\ 0.38476)&(1.5621,\ 0.70148)&(1.0018\ ,0.86675)\\  \hline
   \multirow{2}{30pt}{1000}       & $\beta_v$  \hspace*{2mm} & (1.8920,\ -0.71540)&(2.120, \ 0.62609)&(1.9272,\ 0.64545)\\
                                  & $\beta_s$   \hspace*{2mm} &(1.8724,\ 0.59370)&(1.8750,0.90215)&(1.9035, \ 0.97726) \\  \hline

    \bottomrule

    \end{tabular}
    \label{tb2}
\end{table}
\subsection{Solution of the \lowercase{t}DS equation at ${\mbox{\boldmath$ Re\ \tilde k^2 <0$}} $}
\label{left}

At $Re \, \tilde k^2\ < \ 0$ we use either the Cauchy theorem
 in the domain  where the propagator functions are still analytical or
 the prescription explained below, if there are singularities.
 In  Ref. \cite{ourPRCSingularities} has been  found that  at large values of the
  meson masses $M > 1 $ GeV and at  $ Re ~\tilde k^2<0 $
 the propagator  functions $\sigma_{s,v}(\tilde k^2)$ are not
 longer analytical functions  having an infinite number of pole-like singularities in this region, which in principle
 give rise to numerical problems in solving the tBS equation.
 Nevertheless, as it has also been shown such singularities turn out to be integrable in the tBS equation,
 provided their exact locations and corresponding residues are known. For an analysis of the analytical properties of $\sigma_{s,v}(\tilde k^2)$
 we suggested the following procedure  \cite{ourPRCSingularities}:\\
(i) Choose a relatively large domain within the parabola (\ref{parabola}), enclose it  with a contour and compute the Cauchy integrals
 of $A(\tilde k^2)$, $B(\tilde k^2)$ and the inverse part of the propagator functions $\Pi(\tilde k^2)\equiv (\tilde k A(\tilde k^2))^2+B^2(\tilde k^2)$.
  Vanishing integrals will imply that these functions are analytical within the chosen contour.\\
(ii) Compute  Rouch\'e's  integral\footnote{Rouch\' e's integral of an analytical
complex function $f(z)$ on a closed contour $\gamma$  is defined as
$\frac{1}{2\pi i}\oint\limits_\gamma\frac{f'(z)}{f(z)}d z$.} of the function $\Pi(\tilde k^2)$. Since in the previous item
 $\Pi(\tilde k^2)$ has been found to be  analytical, such an integral, according to Rouch\'e's theorem,  gives exactly the number of its zeros  inside the  contour.\\
(iii) Compute the Cauchy integral of the propagator functions $\sigma_{s,v}(\tilde k^2)$  which,
   if  Rouch\'e's integral of $\Pi(\tilde k^2)$  is found to be   an integer positive
number, clearly must be different from zero. Moreover, it would imply that $\sigma_{s,v}(\tilde k^2)$ have poles inside
the contour and their Cauchy integrals provide the corresponding
 residues, necessary in further applications, see below \\
(iv) Shrink  the area of the contour until the Rouch\'e's integral becomes equal to one
and  continue to squeeze the contour, by keeping the value of the Rouch\'e's integral unchanged,  until
the location of the pole is found with the desired accuracy. \\
The integrals to be calculated are:

  \begin{eqnarray}&&
  \oint\limits_\gamma \left[ \xi^2 A^2(\xi))+B^2(\xi)\right ] d \xi^2 =0,  \label{ucl}\\ &&
  \frac{1}{2\pi i} \oint\limits_\gamma
 \frac{  \left[ \xi^2 A^2(\xi))+B^2(\xi) \right]'_{\xi^2} }{\xi^2 A^2(\xi)+B^2(\xi) }d\xi^2 =N_z,\label{npoles} \\[1mm] &&
  \frac{1}{2\pi i}\oint\limits_\gamma \sigma_{s(v)}(\xi^2) d\xi^2 =
  \sum_i res[\sigma_{s(v)}(\xi_i^2)].
  \label{residv}
 \end{eqnarray}

  We find that  at $M < 1$ GeV all the integrals (\ref{ucl})-(\ref{residv}) are zero, implying
  that the tDS solution $A(\tilde k^2)$ and $B(\tilde k^2)$ and  the propagator functions $\sigma_{s,v}(\tilde k^2)$
  are analytical within the parabola (\ref{parabola}).
  At $M > 1$ GeV   the
  Cauchy integrals for $A(\tilde k^2)$, $B(\tilde k^2)$ and $\Pi(\tilde k^2)$  are still zero, i.e. the inverse propagator
  $\Pi(\tilde k^2)$ is still analytical, while for
  $\sigma_{s,v}(\tilde k^2)$ the Cauchy integrals do not longer vanish. Moreover,  Rouch\' e's integrals of
  $\Pi(\tilde k^2)$  are found to be integer positive numbers which  clearly  indicates that
  $\sigma_{s,v}(\tilde k^2)$ have poles in this region. The number of poles in the initial contour is given by
   the value of  Rouch\' e's integral (\ref{npoles}).

 For numerical calculations it is extremely important  to find, with a good accuracy,
   the position of the poles and the corresponding residues for   $\sigma_{s}(\tilde k^2)$ and   $\sigma_{v}(\tilde k^2)$.
  In such a case, if the  complex valued functions $\sigma_{s,v}(\tilde k^2)$
  have only isolated poles $\tilde k_{0i}^2$ within a certain domain and are analytical along
   its closing contour $\gamma$, they can be represented as
 \begin{equation}
 \sigma_{s,v}(\tilde k^2) = \widetilde\sigma_{s,v}(\tilde k^2) + \sum_i\frac{{\rm res} [\sigma_{s,v}(k^2_{0i})]}{\tilde k^2-k^2_{0i}},
\label{main}
 \end{equation}
 where $\widetilde\sigma_{s,v}(\tilde k^2)$ are analytical functions within
 the considered domain which can be computed as
 \begin{equation}
 \widetilde\sigma_{s,v}(\tilde k^2)
  =\frac{1}{2\pi i}\oint\limits_{\gamma}\frac{\widetilde \sigma_{s,v}(\xi)}{\xi -\tilde k^2} d\xi
 =  \frac{1}{2\pi i}\oint\limits_{\gamma}\frac{\sigma_{s,v}(\xi)}{\xi -\tilde k^2} d\xi .
  \label{f1}
 \end{equation}
 Such a representation~(\ref{main}) of the propagator functions in the presence of pole-like singularities
 allow to avoid numerical problems in calculations of the kernel (\ref{kernel}). The product of two propagator
 functions in~(\ref{kernel}) in the presence of
singularities receives  the form
\begin{eqnarray} &&
\sigma_1(\tilde k_1^2) \ \sigma_2(\tilde k_2^2) =  \widetilde\sigma_1(\tilde k_1^2)
\widetilde\sigma_2(\tilde k_2^2)+\nonumber\\[2mm]&&
+\widetilde\sigma_1(\tilde k_1^2) \sum_j \frac{res[\sigma_2(k^2_{0j})]}{\tilde k_2^2-k^2_{0j}}
+\widetilde\sigma_2(\tilde k_2^2) \sum_i \frac{res[\sigma_1(  k^2_{0i})]}{\tilde k_1^2-k^2_{0i}}
+\sum_{i,j} \frac{res[\sigma_1(k^2_{0i})] res[\sigma_2(k^2_{0j})]}{(\tilde k_2^2-k^2_{0j})(\tilde k_1^2-k^2_{0i})},
\label{doubleres}
\end{eqnarray}
where $k^2_{0i}$ and $k^2_{0j}$ are the positions of the poles of
the propagator functions $\sigma_{1,2}(\tilde k_{1,2}^2)$ of the first and second quark
with the corresponding residues $res[\sigma_{1}(k^2_{0i})]$  and $res[\sigma_{2}(k^2_{0j})]$ in
the integration  domain
of the tBS equation.

  The first term in (\ref{doubleres}) is analytical everywhere within the integration
  domain of tBS and can be computed numerically by
  Eq.~(\ref{f1}), while the last  term in Eq.~(\ref{doubleres}) has already an integrable form which,
  together with the Gegenbauer polynomials from~(\ref{kernel}),  can be easily reduced to a sum of
 integrals like Eq. (\ref{base}), see Appendix. The second and the third terms in~(\ref{doubleres}) are still
 complex functions with singularities. However, the position of these singularities are exactly
 the same as found before, so that  they
  can  be again presented in the form~(\ref{main}) as a sum of analytical function and a pole-like structure and
  reduced to integrals of the type~(\ref{base}).

  Equation (\ref{doubleres}) represents
  the main ingredient of our approach allowing to handle singularities in the tBS kernel~(\ref{kernel}).

\subsection{Location of singularities}\label{poles}
\begin{figure}[!ht]
\includegraphics[scale=0.5 ,angle=0]{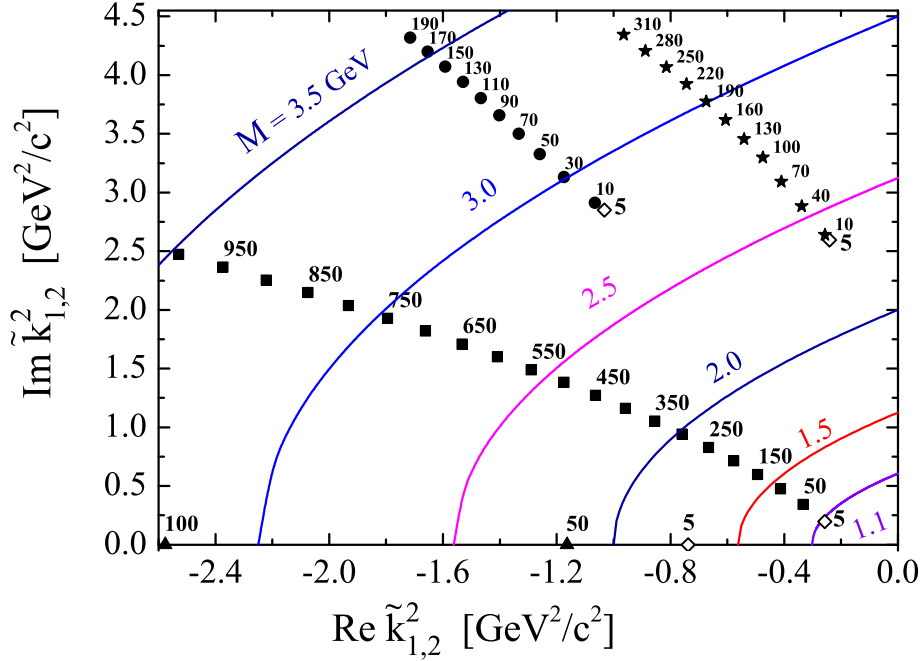}
\caption{(Color online)
  Positions of few first poles (full symbols) in the upper left hemisphere
  of the complex $\tilde k^2$ plane
  for various current quark masses (depicted in MeV).
  The relevant sections of the parabola~(\ref{parabola})
    corresponding to the meson bound-state mass $M$
  are presented by solid curves  for $M=1.1, \, 1.5, \, 2.0,\,  2.5,\, 3.0 $ and $3.5$
  GeV, from right to left.
 The pole positions for a $5$ MeV quark are labeled by open symbols.
}
\label{rectangular}
\end{figure}

We calculated the integrals~(\ref{ucl})-(\ref{residv})
 for different current quark masses in the tDS equation~(\ref{sde}) from
$5$ MeV, the mass corresponding to the light u, d quarks, up to $1$ GeV, which corresponds to the c quark.
Results are presented in Fig. \ref{rectangular}, where the location of first relevant  poles
is depicted for quarks with different current masses $m_q$, the values of which  label the corresponding symbols
in the figure. Also, the portions of the Euclidean space relevant to solve the tBS equation
for bound states $M=1.1,\,\,   1.5,\,\,  2.0,\,\,  2.5,\,\,  3.0$ and $3.5$ GeV from left to   right  are
presented as  domains enclosed by  correspondingly  labeled parabolas (\ref{parabola}).

   Figure~\ref{rectangular} allows, even prior to solve the tBS, for a rather general
   analysis of analytical properties of the meson bound states.
  For instance, it is seen that \\
  (i)  for meson masses $M<1$ GeV there are no singularities in the tBS kernel,\\
  (ii) the first two self-conjugated poles appear at $M\sim 1$ GeV and belong to light
       u, d quaks (the open diamonds in the figure),\\
  (iii) the pole-structure of s quark propagators ($m_q\sim 115$)  MeV starts at $M > 1$ GeV and up to
    $M\sim 2.5$ GeV there are only two, self-conjugated  poles. The third and forth  poles for strange
    quarks are enclosed in parabolas $2.5 \, {\rm GeV}\, <\, M< 3.0\,  {\rm GeV}$ and
    $3.0 \, {\rm GeV}\, <\, M< 3.5\,  {\rm GeV}$,
    respectively, \\
  iv) for the charmed quarks the singularity are located  near $M\sim 3.5 $ GeV.\\
 From this quick analysis one infers that calculations of mesons with masses  $M\leq 1.1$ GeV
  (such as $\pi$, $\rho$, $K$, $\phi$ mesons) do not encounter difficulties related to
  singularities.
  For mesons with at least one light quark and  $M\ge 1.1$ GeV
  (such as $D$ mesons, excited states of pions and kaons etc.) the tBS kernel contains propagator functions
  with singularities, which ought  to be treated accurately,  as explained above.

 \begin{table}[!ht]
 \caption{ The position and residues of first few poles of propagator functions for u, d quarks.}
    \begin{tabular}{clll} \hline
    i & \hspace{2mm}positions                          & \hspace{2mm} res[$\sigma_s$]                        &\hspace{2mm}res[$\sigma_v$]  \\ \hline
    1  &\hspace{2mm} (-0.2585,\ 0.1958)    & \hspace{2mm} (-1.6130 $10^{-2}$,\ -0.5109)&\hspace{2mm} (0.2589,\ - 0.8596)\\
    2  &\hspace{2mm} (-0.2409,\ 2.5947)    &\hspace{2mm}   (4.0336 $10^{-2}$,   0.1003)&\hspace{2mm}  (2.3384 $10^{-2}$,\ 6.2750 $10^{-2}$)  \\
    3  &\hspace{2mm} (-0.7382,\ 0.0)      &\hspace{2mm}   (6.8860 $10^{-2}$, \ 0.0)  & \hspace{2mm} (-8.0147 $10^{-2}$,\  0.0)  \\
    4  &\hspace{2mm} (-1.0415,\ 2.8535)    &\hspace{2mm}   (-0.05,\,  0.076)  &\hspace{2mm} (0.0014,\  -0.052)  \\
    \hline \end{tabular}
\label{tbu}
\end{table}

For completeness, in Table~\ref{tbu} we present the location of poles and their residue
for the u, d quark, relevant to most calculations of mesons at $1<M< 3.5$ GeV.
Also it is worth to point out that the propagator functions for the  s  quark ($m_q=115\, MeV$) possess only a self conjugated
pole in the vicinity of the considered parabolas, located at $k_{01}^2=(-0.436 \pm  0.5131 i) {\rm (GeV/c)}^2$
with residues  ${\rm res}[\sigma_s]=(9.05\, 10^{-3} \mp  0.491 i)$ GeV
and ${\rm res}[\sigma_v]=0.261 \mp  0.538 i $ for $\sigma_s$ and $\sigma_v$, respectively.
The second pole, located at $k_{02}^2=(-0.507 \pm 3.35 i)~{\rm (GeV/c)}^2$ (with the respective residues
${\rm res}[\sigma_s]=(5.5\, 10^{-2} \pm  0.10 i)$ GeV and
  ${\rm res}[\sigma_v]=1.34, 10^{-2} \mp  6.12, 10^{-2}i $), is located already too far from the
corresponding parabola for strange mesons and, consequently is irrelevant in numerical calculations.

\section{Numerical methods}\label{NumMetods}
The performed analysis of quark propagators allows one
to calculate the kernel of the tBS equation in the whole  region   of
the complex Euclidean space relevant for the tBS equation.
In   numerical calculations we form the skeletons of approximate
solutions and kernels by using the Gaussian   method of computing  integrals
and by restricting the infinite sum over $n$ in   Eqs. (\ref{er})-(\ref{eqnf})   by
a finite value $M_\mathrm{max}$.
The Gaussian quadrature formula assures a rather good convergence of the
numerical procedure  and provides the
sought solution in the Gaussian nodes which are spread rather uniformly in the
 interval $0\le \tilde p < \infty$. In order to have the solution in  detail
at moderate values of $\tilde p$, which is the interval of the actual physical interests,
one usually redistributes the Gaussian mesh making
the nodes  more dense at low values of $\tilde p$.
To this end one  applies an appropriate mapping of the Gaussian mesh
by  changing  of variables as, e.g.  in Ref.~\cite{dorByer}.
 The resulting system   of linear equations reads then as
\begin{eqnarray}
X= \, SX,\label{syst}
\end{eqnarray}
where  the vector
\begin{eqnarray}
X^T=\left ([\{\varphi_1^n({\tilde k}_i)\}_{i=1}^{N_G}]_{n=1}^{M_\mathrm{max}},
[\{\varphi_2^n({\tilde k}_i)\}_{i=1}^{N_G}]_{n=1}^{M_\mathrm{max}},
\ldots,
[\{\varphi_{\alpha}^m({\tilde k}_i)\}_{i=1}^{N_G}]_{n=1}^{M_{max}}\right )
\label{vector}
\end{eqnarray}
represents the sought solution in the form of a group of sets of
partial wave components $\varphi_\alpha^n$,
  specified on the integration mesh of the order $N_G$.
 The matrix $S$ is determined by  the corresponding  partial kernels (\ref{kernel}),
the Gaussian weights
and the  Jacobian of the mapping and is of the
 $N\times N$ dimension, where $N=\alpha_\mathrm{max}\times M_\mathrm{max}\times N_G$.
Since the system of equations (\ref{syst}) is homogeneous,
the  eigenvalues  of the bound state with mass $M$
is obtained from the condition
$\det(S-1)=0$. Then,
the partial components $\varphi_\alpha^n$ are
found by solving  numerically
the system (\ref{syst}) at this bound-state mass $M$.

We use a combined method of finding the solution $X$.
 First, the Gauss-Jordan elimination and pivoting method
  involving the choice of the leading element
  is applied.
  Then the obtained solution is used as a trial input into an
   iteration procedure to find (after $ 5-10 $ iterations)
  more refined results.
  \subsection{Pseudo-scalar meson ground  states results}\label{ps:mesons}

\begin{figure} \begin{center}
\includegraphics[width=0.9\textwidth]{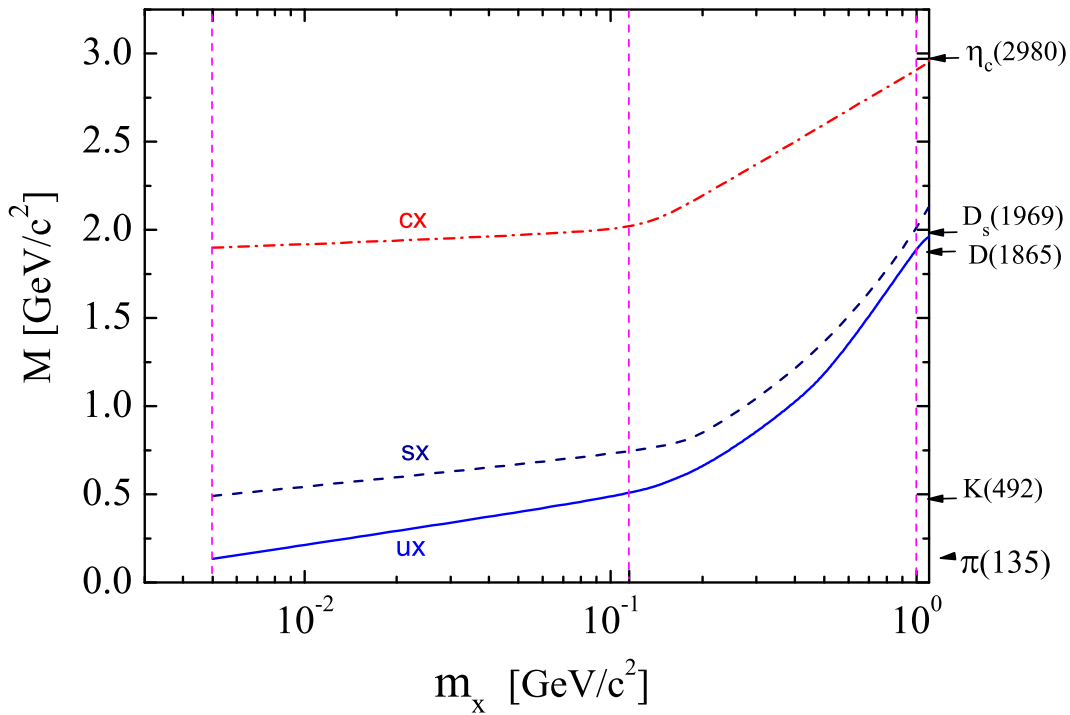}\end{center}
\caption{(Color online) The bound-state masses (ground states)
of a system $qq_x$ as a function of the bare  mass
${m}_x$, for
${m}_q=0.005$ GeV (the curve labeled as "$ux$")
or  ${m}_q=0.115$ GeV (the curve labeled as "$sx$") or  ${m}_q = 1.0$ GeV (the curve labeled as "$cx$").
The effective parameters of the vertex-gluon kernel~(\ref{phenvf})  are  $\omega = 0.5$ GeV and $D=16$ GeV$^{-2}$.
The masses of the pseudoscalar $\pi$, $K$, $D$, $D_s$ and $\eta_c$ mesons according
to \cite{PDG} are indicated at the right side.
 The vertical dashed lines mark the selected bare quark masses for  u, s, and c.}
\label{fig3}
\end{figure}

As an example of our numerical study we exhibit in Fig.~\ref{fig3}
the energy of the lowest bound states of a hypothetical meson $qq_x$ consisting of one given quark
 $q$  with the mass known from the tDS equation, bound with
a second quark $q_x$ for which the input bare mass  ${m}_x$  is let to  vary arbitrarily.
The corresponding effective parameters have been chosen as mentioned above
and the bare
masses for $q$ correspond to u, d, s and c quarks,
$q$=u (with $ {m}_u = 0.005$ GeV, solid line), $q$=s (with $ {m}_s = 0.115$ GeV, dashed line)
and $q$=c (with $ {m}_c = 1.0$ GeV, dot-dashed line).
This figure illustrates   the whole mass spectrum of pseudoscalar mesons with masses up to $3$~GeV.
In obtaining  stable results for  masses above $1$ GeV, the pole structure of the propagators
of the corresponding $q$ and $q_x$ quark have been treated as explained above, cf.
Eq.~(\ref{doubleres}) and  Fig.~\ref{rectangular}.

As mentioned, Fig.~\ref{fig3} encodes results of ground-state masses for
known pseudo-scalar mesons below 3 GeV. So,  if
 the $q_x$ quark corresponds to a  c  quark, then at the intersection of the vertical line
${m}_x=m_c$ ($\approx 1 \,\, GeV/c^2$) with the "ux" curve one obtains the $D$ meson (with the quark contents   uc),
with the "sx" curve the $D_s$ meson and with the "cx" curve  the $\eta_c$ meson, respectively.
It is worth noting that the "ux" curve crosses the
${m}_u=0.115$ GeV line roughly at the same value of $M$ as the "sx" curve crosses
the ${m}_u=0.005$ GeV line,
thus providing a check  of consistency of the approach and, at the same time,
describing correctly the lowest pseudoscalar $us$ state corresponding to the K meson.
It can be seen that even without a fine tuning of ${m}_{s,c}$ the meson mass spectrum
is reproduced fairly well:  135 MeV ($\pi^0$ meson),
 497~MeV ($K$ meson), 1870 MeV ($D^\pm $ meson), 1970 MeV ($D_s^\pm $ meson)
 and 2980 MeV  ($\eta_c$ meson).

 \subsection{ Vector meson states  }\label{vvted}
As mentioned above, the minor difference  in
calculations of pseudo scalar and vector meson masses consists  in the fact that the   basis of the
spin angular harmonics~(\ref{nharmd}) in the latter case has eight
components instead of four in the former case.  The general structures of the
tBS kernel $S_{\alpha\beta}$ ($\alpha,\beta =1\ldots 8$), Eq. (\ref{kernel}), and of $A_{\alpha\beta}$, Eq. (\ref{Aab}),
remain  the same. The propagator functions $\sigma_{s,v}$, being the solutions of
the tDS equation, do not depend on the meson spin and  are  as before. That means that the present approach
allows to perform, in the same manner, calculations of mass-spectra of mesons of any spin (scalar, pseudo-scalar, vector etc.), cf.~\cite{lastFischer}.
Here it is worth emphasising  that, as one can infer from  Fig.~\ref{rectangular},
the propagator functions $\sigma_{s,v}$ do not contain
singularities for  vector mesons with ground-state masses up to
$3.5$~GeV (such as   $\rho$, $\phi$,  $J/\Psi$ mesons). Consequently, all numerical
calculations can be safely performed  as in the previous approaches,
cf.~\cite{Blank:2011ha,ourFB,lastFischer}, without
accounting for singularities.
Likewise in the case of pseudo-scalar mesons,  the performed analysis
of the vector meson spectra shows that,  results of numerical  calculations
are in an amazing    good agreement with experimental data, cf.~\cite{Blank:2011ha} .
 \subsection{ Excited meson states  }\label{excited}

\begin{figure} \begin{center}
\includegraphics[width=0.9\textwidth]{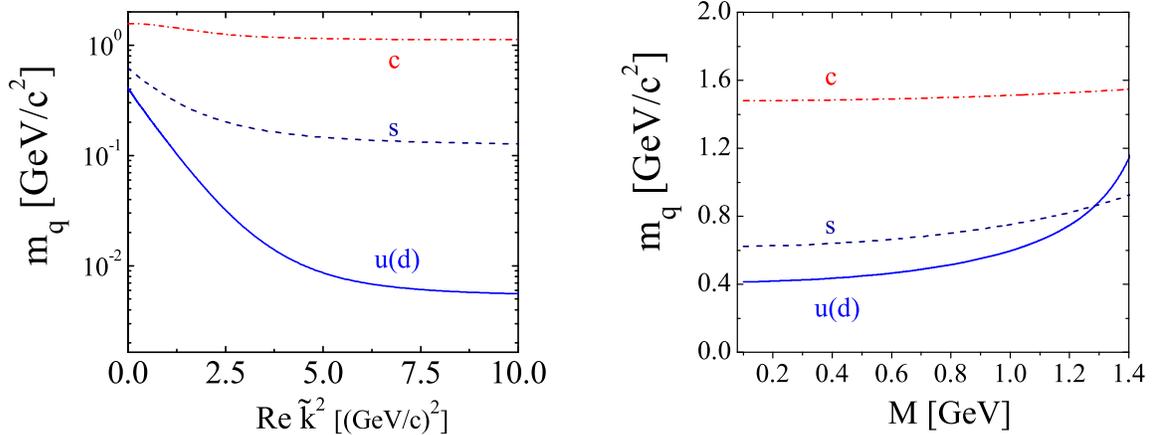}\end{center}
\caption{(Color online) Dynamically dressed quark  mass $m_q(\tilde k^2) =B(\tilde k^2)/A(\tilde k^2)$
from solutions of the Dyson-Schwinger
equation  along the real axis, i.e. for  $Im \tilde k^2 =0$.
The left panel illustrates   the dependence  on  $Re k^2 > 0$,
i.e.  the behaviour of the solution of the tDS equation in real Euclidean space.
  In the right
panel, the  quark masses $m_q$ are calculated  at the parabola vertex, $Re k^2 =-M^2/4$, and
 displayed as a function  of  $M$. A comparison of the two panels
  illustrates   the dynamical  mass increase in the   complex Euclidean plane. Solid curves are
  for the light
u, d quarks, dashed curves  for s quarks, and dot-dashed curves represent c quarks.}
\label{fig3Mas}
\end{figure}

Within the present approach a description of  radial excitations of mesons
is straightforward. It suffices
to find the next zeros of the determinant $det(S-1)$ of the system~(\ref{syst}).
It is worth noting that, in an analogy with
the tBS equation for constituent quarks  with constant masses,  one would, at first glance, expect
that the next zero  of the determinant is to be searched for in a region of $M$
which dos not exceed the maximum value  of masses of two dressed quarks.
The tDS solution provides the corresponding maximum
around $400$ MeV for the light u, d quarks, $600 $ MeV for s quarks and
 $1550$ Mev for c quarks, see  Fig. \ref{fig3Mas}, left panel.
 Correspondingly,   one would expect that   $u\bar u$  excited
 states cannot have a  mass  larger than $800 $ MeV. Analogous restrictions can be
 "deduced" for bound states with s and c quarks.

  However,  in the  non-Euclidean  domain, which corresponds to $\tilde k_{1,2}^2\simeq -M^2/4$,
   the dynamical quark masses $m_q(\tilde k^2) =B(\tilde k^2)/A(\tilde k^2)$ contributing to
   the tBS equation can strongly increase
    in dependence on $M$. This situation is illustrated in Fig. \ref{fig3Mas}, right panel,
    where the dressed quark masses are depicted at the parabola vertex
    $Re~\tilde k^2=-M^2/4$ as a
    function of $M$.
   The corresponding masses $m_q(\tilde k^2)$ are  larger in comparison
   with the net solution of the tDS equation, left panel.
     Hence, the Dyson-Schwinger equation allows to understand
   the formation mechanism of excited states which, from the constituent quark model point of view,
    can not be even predicted a priori. It should be emphasized  that
    the quark masses $m_q$ in the complex Euclidean
    space have also singularities  related to  $A(\tilde k^2)=0$. If, in solving the tBS equation,
     instead of
    propagator functions $\sigma_{s,v}$  one deals with the quark mass as independent variable,
    an analysis of the  singularities must be performed in detail  prior  solving the tBS
    and to find an adequate implementation  in the numerical algorithm.
    It turns out that such an analysis becomes   more involved than the
     one performed for the propagator functions  presented above.

   Coming back to calculations of the radially excited states, we mention that the determinant $det(S-1)$ revealed
    its  first excited $u\bar u$ state at $M \approx 1080$ MeV, i.e. significantly  above the maximum mass
    delivered by
   the Dyson-Schwinger equation alone at $Re ~\tilde k^2>0$.
   The third  zero of the determinant for   $u\bar u$ states has been found around $1300$ MeV, corresponding to
   $\pi$(1300)~\cite{PDG}.
   Analogously,  for the $ c \bar u$ system, the first excited state is found to be around  2530 MeV, which is in a good
   agreement   with data, cf.~\cite{PDG}. Similar results have been obtained also by other groups, see e.g.
    Ref.~\cite{krassnigg1}.

   \subsection{Exhausting method}\label{exhaust}
The   above  method of finding zeros of the determinant of the integral kernel
is rather universal, provided the analytical properties of the kernel itself are known.
However, the method becomes quite cumbersome
if one tries to increase the accuracy of calculations by increasing the number $N_G$ of the Gaussian mesh
and the number $M_{max}$  of terms in the Gegenbauer decomposition. In this case, the dimension of the
determinants drastically increases and the method  of solving it for zeros becomes a challenging procedure.
Another   approach is to use  an iteration method for solving the corresponding equations.
As  mentioned above,  a finite  kernel of  a Fredholm type equation
  has a discrete   and real spectrum with a nondecreasing sequence of   eigenvalues $\lambda$.
   Moreover, it can be proven  that, if such an equation is solved by iterations,
it converges to the lowest value of the spectrum, i.e. to the ground state of the equation.
It means that  one can solve the BS equation as an eigenvalue problem,
\begin{eqnarray}
X= \, \lambda (M) SX \label{syst1},
\end{eqnarray}
for the eigenvalue $\lambda (M)$ as a function of the bound state mass. Then, the sought solution
 $M^{g.s.}$
 can be found at $\lambda (M^{g.s.})=1$. Then the kernel can be modified
 for the use of a iteration scheme for exited states. To this end,  once the first eigenvalue is found, one
 constructs from this solution and from the old kernel  a new one by subtracting from the previous kernel
 the contribution of its ground state.
 It is possible to do it in such a way that the new kernel  will have the same spectrum of eigenvalues as the
 previous one except the eigenvalue already found.
 Obviously, for the new kernel the  iteration method provides its ground state, which  actually is the
 first excited state of the previous kernel. By continuing this procedure for the next eigenvalues one can
 find all the desired excited states. Such a method
 is known as the exhausting (depletion) method and was reported in some details  in Ref.~\cite{ourFB1}.
 It seems that a similar method has been recently employed in Ref.~\cite{lastFischer}.

 This method holds only if the integral kernel is finite. In the case
 when the kernel $S$ contains singularities through
 the propagator functions, one has to handle them accurately as to obtain the desired finite $S$.
 It should be noted  that often in numerical calculations of the kernel (\ref{kernel})
 the singularities are "overlooked" by not too large integration meshes  used to
evaluate the  two dimensional integrals in  (\ref{kernel}). It means that in replacing the continuum
Euclidean domain of integration (enclosed by the corresponding parabolas) by discrete meshes, one can "jump" over
singularities and perform calculations without any troubles. However, increasing the density of
mesh points one can approach closely the singularities and, consequently, the stability of
solution can be lost.

\begin{figure}
\includegraphics[width=0.9\textwidth]{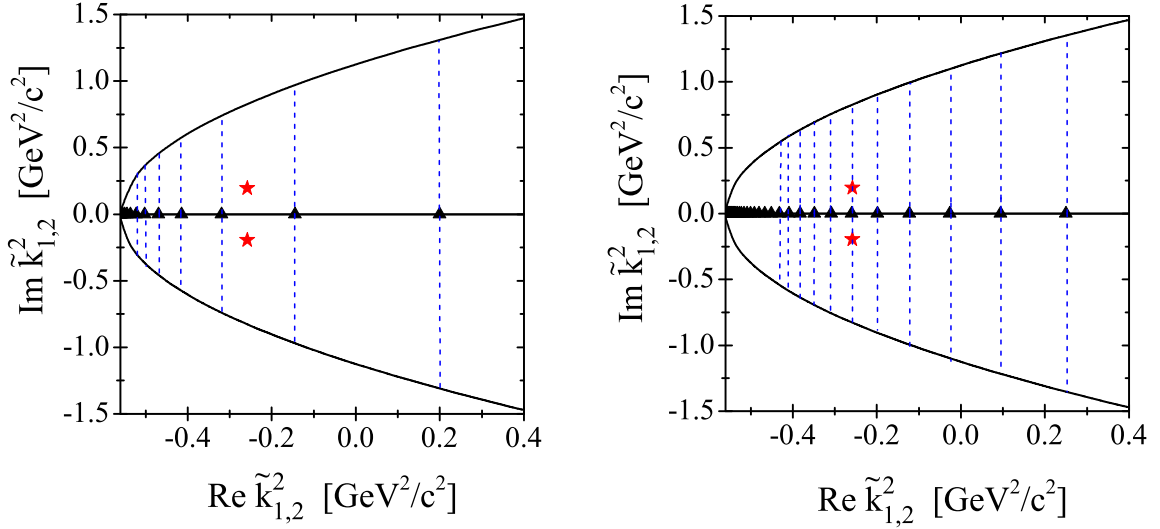}
\caption{(Color online) The interaction
 domain of the tBS equation over the complex $\tilde k_{1,2}^2$ plane which is the surface
 of parabola
$Im~\tilde k_{1,2}^2 =\pm M \sqrt{M^2/4+Re~\tilde k_{1,2}^2}$ for $M=1.5$ GeV. The two self-conjugated
poles for u, d quarks are represented by stars.
 The full
triangle symbols in left and right panels represent the Gaussian
mesh for $Re~\tilde k_{1,2}^2$ with 32 (left)  and 96 (right) nodes, respectively,
 for the $d\tilde k$ integration which
actually extends to $+\infty$. For each value of the node
$Re~\tilde k^2_i$ the angular integration $d\cos\chi_k$ goes along the   vertical dashed lines
(only a few ones are exhibited) corresponding  to $-1\le \cos\chi_k\le +1$. }
\label{illustration}
\end{figure}
In Fig.~\ref{illustration}
we illustrate such a case.
If one uses, e.g.,  Gaussian meshes $(j,\lambda) $
to integrate over $\tilde k$ and $\chi_k$ then in the tBS the corresponding
values of momenta of quarks will be
$ \tilde k^2_{1,2(j,\lambda)} = -M^2/4 +k_j^2 \pm i M\tilde k_j \cos\chi_{k,\lambda}$.
In Fig.~\ref{illustration} these nodes are depicted as $Re~ \tilde k_{1,2}^2(j)=-M^2/4 +k_j^2$
by  full triangle symbols  and,
at  each $Re ~\tilde k_{1,2}^2(j)$ the angular  integration is depicted by vertical dotted lines. In the picture
we adopted $M=1.5$ GeV. The left panel corresponds to a 32 nodes Gaussian mesh for  $Re ~\tilde k_{1,2}^2$, while
the right panel corresponds to a 96 nodes Gaussian mesh.
As mentioned above, in  our calculations we use  a mapping procedure $k_j=k_0\dfrac{1+x_j}{1-x_j}$ with $k_0=0.1$,
where $x_j$ are the Gaussian nodes defined from $-1$ to $+1$.
 For illustration,  the pole position of a $5$ MeV quark is depicted by full stars.
 It can be inferred from the picture that
 in a numerical procedure all
 troubles can occur when the corresponding vertical line (angular integration)
 crosses the poles. At low
 values of Gaussian mesh the poles remain untouched (left panel)
 and the results could be finite without any additional
 treatment of the kernel (\ref{kernel}).
 By increasing the Gaussian mesh the probability to meet the poles increases (right panel).
 In this case the tBS equation cannot be reliably solved numerically.

 In principle, to avoid these problems one can use also the so-called "extrapolation" method, by calculating the
 dependence $\lambda (M)$ for masses $M < 1$GeV, for which the kernel is an analytical function,
 then to extrapolate $\lambda (M)$ in the region with poles, see Ref.~\cite{lastFischer}.
  This method, if used not too far beyond the region of analyticity
  of the corresponding propagators,  provides also rather good agreements with experimental data.

 \subsection{Equal quark masses, normalization condition}\label{equalMass}
A particular situation occurs in case of mesons with $M > 1$ GeV  formed by quarks of equal masses,
e.g. in calculations of the excited states of pions above $1$ GeV or in calculating the normalization
of the tBS partial amplitudes with at least one quark with poles,
i.e. in calculations one meets   products of two identical propagators with singularities.
In such a case one has $k_1^2=k_2^{*2}$ and,  due to self-conjugated nature
of the propagator functions  in Eq.~(\ref{doubleres}),
one has also $\sigma_2(k_2^2)=\sigma_1^*(k_1^2)$. Moreover,
 the poles coincide for both propagators and in the sums over poles in (\ref{doubleres})
 one also encounters the situation with $k_{0i}^2= k_{0j}^{*2}$. In this case, i.e.
 the $Re ~\tilde k_{1,2}^2$
 exactly   coincides with the $Re ~k_{0i}^2$ in the integrand (see right panel in Fig.~\ref{illustration}),
the angular integration reduces to (see Appendix, Eq.~(\ref{zs})).
 \begin{eqnarray}
 {\cal I}^\lambda_{mn}(y)\sim \int\limits_{-1}^1 d\xi (1-\xi^2)^{\lambda-1/2}  \frac{ G_m(\xi) G_n(\xi)}{(\xi-y)^2},\label{y}
 \end{eqnarray}
 where $y$ is purely real and depends on the position of the pole, $y=Im ~k_{0i}^2/M\tilde k$.
 Since the pole is supposed to be inside the parabola one always has
 $|y|<1  $.  For such values of $y$, the integral (\ref{y}) is not accessible in
quadratures. Obviously, this  situation is accidental and occurs only because of our specific
 choice of the Jacobi coordinates $k_1$ and $k_2$ when both quarks carry equal portions of the
 total momentum $P$ of the bound state. The problem  can be solved by redistributing the total momentum $P$
  between quarks as
 \begin{eqnarray}
 k_1=\eta P + k, \qquad k_2=(1-\eta) P - k
 \end{eqnarray}
 with $\eta\neq 0.5$. In this case, the momenta of quarks are no  longer mutually
 complex conjugated, and the corresponding expression becomes again integrable.
  In our calculations of excited states of the pion we slightly changed $\eta$ away
  from $\eta=0.5$ by
$\sim 10\%$ which allows us to perform safely the necessary calculations.
In principle, the solution of the BS equation in the ladder approximation
 must be independent of $\eta$, see e.g. Ref. \cite{itzikson}, which,  in the rainbow approximation,
 was numerically confirmed  in several papers cf.~\cite{rob-1,Maris:1999nt,souchlas}.
  It should be noted, however, that in
 concrete calculations where  the Gegenbauer expansion is truncated at some value
 $M_{max}$ the numerical
 solution may depend on the choice of $\eta$. In our calculations
 we investigated the sensitivity of numerical solutions on $M_{max}$ and found
 that for $\eta=0.5$ the  solution of the  tBS equation is stable  already at $ M_{max}=4-5$.
 It remains stable also  if we change $\eta$ by $\sim $ ( 5-10)\% . Larger deviations of $\eta$
 will require a reanalysis of the dependence
 on  $M_{max}$.

A few comments about the normalization of the tBS amplitude are in order here.
 Since the BS equation is a homogeneous equation,
the solution for the amplitude  must be additionally normalized. The normalization condition in the  general case
has been derived in Ref.~\cite{nakanishi}.  For our case it reads \cite{tandy1,Alkofer}
\begin{eqnarray}
\frac32 \frac{\partial}{\partial P_\mu}  Tr \left[
\int\frac{d^4 k}{(2\pi)^4} \bar \Gamma(-P,k) S(k_1)\Gamma(P,k)S(-k_2)
 \right] = 2P_\mu .
 \label{norm}
 \end{eqnarray}
In the ladder approximation, due to translation invariance  the tBS amplitude, $\Gamma(P,k)$ does not depend
 on the total momentum
$P$ and often the derivative ${\partial}/{\partial P_\mu}$ is moved
 inside the integral acting only on $S(k_{1,2})$. Such an operation is mathematically correct only
 if the integral is absolutely convergent. If so, then the derivative
$ {\partial}/{\partial P_\mu} =\eta  \, {\partial}/{\partial k_{1\mu}} $ acting, e.g.,
 on the first propagator
$S(k_1)$,
leads to  an expression of the form
\begin{eqnarray}
{\partial S(k_1)}/{\partial k_{1\mu}} = -S(k_1) [i\gamma_\mu A(k_1) +A'(k_1)+B'(k_1)] S(k_1),
\end{eqnarray}
i.e. after calculations of the corresponding  traces, the remaining integral will contain
two identical propagators. Calculations of such integrals are cumbersome, but straightforward.

 In the case of singularities, if one still interchanges the derivative with integration,
the remaining integral after traces  lead again to divergences of the type (\ref{y}).
In this case, since the two propagators refer to the same quark,
 even the use of $\eta \neq 0.5$ does not cure the problem. It merely implies that
one cannot exchange derivatives and integrations if the integral is not absolutely convergent.
To normalize correctly the tBS amplitude in the presence of poles in propagators, one must first
calculate the corresponding traces in~(\ref{norm}) as a function of $P_\mu$ and  then evaluate the
derivatives numerically.

\section{Summary}\label{summ}
\label{summary}
 We analyse  the truncated Dyson-Schwinger (tDS) and Bethe-Salpeter (tBS) equations
 in the Euclidean complex momentum domain  which is determined by the
 mass $M$ of mesons as   quark-antiquark  bound states.  Within the ladder rainbow truncation,
   only the infrared term in the combined effective vertex-gluon kernel is retained.
 The locations of singularities of the propagator functions and their residues
 are determined  with   high accuracy  in the whole region relevant to describe
 mesons with energy (masses) up to $M\leq 3.5$ GeV.
 We propose a method of separating
 the analytical part and the pole structure in the propagators   to be
 further implemented easily  in numerical algorithms. It is demonstrated that the part with
 singularities can be integrated explicitly, avoiding in such a way
 difficulties in handling numerically singular quantities. The proposed method
 has been applied to calculate the mass spectra of  pseudo-scalar mesons with
 $M\leq 3.5$ GeV with and without singularities in the propagator functions.
 We obtain  a good agreement with experimental data.

 The performed analysis is aimed at elaborating  adequate
 numerical algorithms to solve the BS equation in presence of singularities and to investigate the properties of
 mesons, such as the open charm $D$ mesons, related directly to  physical programmes envisaged, e.g. at FAIR.
 Then, as our ultimate goal, the performed analysis is to be used as
a base line for investigations of mesons at finite
temperatures and baryon densities.

\section*{Acknowledgments}
This work was supported in parts by the Heisenberg - Landau program
of the JINR - FRG collaboration, GSI-FE and BMBF O5P12CRGH1. LPK appreciates the warm hospitality at the
Helmholtz Centre Dresden-Rossendorf. The authors gratefully acknowledge
discussions with  M. Lutz, M. Viebach, R. Williams T. Hilger and C. Fischer and
thank T.E. Cowan for the support of our research project.

\appendix
\section{Useful relations}

For the product of two propagator functions in (\ref{kernel})
the pole part  (\ref{doubleres})   can  be rewritten as

\begin{eqnarray}&&
\left[\frac{1}{\tilde k_1^2-k_{0i}^2}\right]\left[\frac{1}{\tilde k_2^2-k_{0j}^2}\right]=
\left[\frac{1}{Re \tilde k_1^2 +iM \tilde k\xi-k_{0i}^2} \right]
\left[ \frac{1}{Re \tilde k_2^2 -iM \tilde k\xi-k_{0j}^2}\right]
=\nonumber\\&& =
\frac{1}{(M \tilde k)^2 } \Big[\frac{1}{ \xi- i\left( \tilde Re k_1^2-k_{0i}^2\right)/M\tilde k }\Big ]\cdot
\Big[\frac{1}{ \xi- i\left( -Re \tilde k_2^2+k_{0j}^2\right)/M\tilde k }\Big ]=\nonumber\\ && =
\frac{1}{\Delta z_{ij} M \tilde k}\left[ \frac{1}{\xi -iz_i^{(1)}}-\frac{1}{\xi-iz_j^{(2)}}\right]
\label(a1)
\end{eqnarray}
where
\begin{eqnarray}
Re ~\tilde k_1^2=Re ~\tilde k_2^2=-M^2/4-\tilde k^2, \quad  z_{i,j}^{(1,2)}=\dfrac{\pm Re  \tilde k^2_{1,2}\mp k_{0i,j}^2}{M\tilde k},\quad {\rm and}\
\Delta z_{ij}=z_i^{(1)}-z_j^{(2)}.
\label{zs}
\end{eqnarray}
Then the  main integral in the tBS kernel (\ref{kernel}) with  pole-like  singularities is

\begin{eqnarray}
{\cal I}_{mn}^\lambda(z)=\int \limits_{-1}^1 d\xi (1-\xi^2)^{\lambda-\frac 12} G^{\lambda}_m(\xi)G^{\lambda}_n(\xi)
\frac {1}{\xi-iz} = \nonumber  \\
 -\frac {2\sqrt{\pi}}{\Gamma(\lambda)} \left(\frac 12\right)^{\lambda-\frac 12}
e^{(\frac 12 -\lambda)i\pi}(-z^2-1)^{\frac{2\lambda-1}{4}} G^{\lambda}_{min}(iz)
Q^{\lambda-\frac 12}_{max+\lambda-\frac 12}(iz), \label{a2}
\end{eqnarray}
where
$G^{\lambda}_m(x)$ are the   Gegenbauer polynomials obeying  the following recurrent relations
\begin{eqnarray} &&
2\lambda(1-\xi^2) G^{\lambda+1}_{n-1}(\xi)=(2\lambda+n-1)G^{\lambda}_{n-1}-n \xi
G^{\lambda}_n(\xi) \\ &&
(n+1)G^{\lambda}_{n+1}(\xi)=2(n+\lambda)\xi
G^{\lambda}_n-(n+2\lambda-1)G^{\lambda}_{n-1}(\xi),
\end{eqnarray}
$Q^{\lambda}_m(x)$ are the Legendre functions
of the second kind and  $min=min(m,n)$, $max=max(m,n)$.

  Some useful relations with Legendre functions related to computations of our integral are
\begin{eqnarray} &&
Q^{1/2}_{n+1/2}(iz)(-z^2-1)^{1/4}=-i^n\sqrt{\frac {\pi}{2}} Z^{n+1},\nonumber \\ &&
Q^{3/2}_{n+1/2}(iz)(-z^2-1)^{3/4}=-i^{n+1}\sqrt{\frac {\pi}{2}}\left( n z
Z^{n+1}+(n+1)Z^{n}\right),  \nonumber \\ &&
Q^{3/2}_{n-1/2}(iz)(-z^2-1)^{3/4}=-i^n \sqrt{\frac {\pi}{2}} \left( (n-1) z
Z^{n}+nZ^{n-1}\right),   \\ &&
Q^{5/2}_{n+1/2}(iz)(-z^2-1)^{5/4}=i^n\sqrt{\frac {\pi}{2}}
\left( n(n-1)z^2
Z^{n+1}+(n-1)(2n+3)zZ^n+n(n+2)Z^{n-1} \right)\nonumber, \\&&
Z=(z-\sqrt{z^2+1})=\frac {-1}{z+\sqrt{z^2+1}}.
\nonumber \end{eqnarray}

\end{document}